\documentclass[12pt,preprint]{aastex}

\begin{document}

\title{The Optical Orbital Light Curve of the Low-Mass\\ 
       X-Ray Binary V1408~Aql (= 4U~1957+115)}
 
\author{Amanda J. Bayless\altaffilmark{1,2}, 
        Edward L. Robinson\altaffilmark{1}, 
        Paul A. Mason\altaffilmark{3,4},
        and
        Paul Robertson\altaffilmark{1}}

\altaffiltext{1}{Department of Astronomy, University of Texas at Austin,
       1 University Station, C1400, Austin, TX 78712}
\altaffiltext{2}{Space Science and Engineering Division,
       Southwest Research Institute,
       6220 Culebra Road, San Antonio, TX 78228}
\altaffiltext{3}{Department of Physics, University of Texas at El Paso,
                 El Paso, TX 79968}
\altaffiltext{4}{Department of Mathematics and Physical Science, 
                 New Mexico State University - DACC, Las Cruces, NM 88003}

\begin{abstract}
V1408 Aql (= 4U 1957+115) is a low-mass X-ray binary with an orbital
period near 9.3 hr, whose compact star is a black hole candidate. The
system shows a large-amplitude orbital photometric modulation at optical wavelengths.
We have obtained new optical photometry of V1408 Aql from which we
derive the orbital light curve and an improved orbital ephemeris. We show that
the orbital light curve can be reproduced by a model in which the accretion disk
around the compact star is thin, axisymmetric, and uneclipsed. The secondary
star is heated by X-rays from the compact star and the accretion disk. The orbital
modulation is produced entirely by the changing aspect of the irradiated
secondary star with orbital phase. Because the system does not eclipse, the fits
of the model light curves are insensitive to the orbital parameters, allowing a
wide range of orbital inclinations and mass ratios.

\end{abstract}


\section{Introduction}
Low-mass X-ray binaries (LMXBs) are close binary stars containing
a black hole or neutron star surrounded by an accretion disk 
fueled by mass transferred from a low-mass companion. 
V1408~Aql is an LMXB with
an orbital period of $P =9.329 \pm 0.011$~h, derived by \citet{tho87}
from a roughly-sinusoidal modulation he found in its optical light curve.  
\citet{tho87} attributed the modulation to
the varying aspect of the secondary star, which is heated by 
flux from the primary star and its accretion disk.
\citet{hak99} obtained two nights of $UBVRI$ photometry, covering 
one orbital period in total, and also saw an orbital variation, but
the light curve was more asymmetric than sinusoidal.
Noting that V1408~Aql became bluer when fainter, \citet{hak99}
interpreted the minimum in the light curve as a grazing eclipse 
of a cool, red rim on the accretion disk by the secondary star.  
In their model the ratio of the rim height to disk
radius is large, $H/R\sim0.2-0.5$, and
the orbital modulation outside eclipse is caused by
variations in the height of the rim, which obscures more or less
of the brighter inner disk.    
Because it invokes an eclipse, the model implies a high 
orbital inclination, the best fitting model having an inclination 
near $75^{\circ}$.

4U~1957+115, the X-ray counterpart of V1408~Aql, is a 
persistent X-ray source, showing neither large transient increases 
in brightness nor faint quiescent states.
Its X-ray flux does, however, vary on all time scales from
10~Hz (1-2 percent rms amplitude) to years (by a factor of four)
\citep{wij02, rus10}.
The orbital period has not been detected at X-ray wavelengths
but \citet{lev06} reported a ``marginal detection'' of a periodicity at 
$9.3175\pm 0.0005$ h, perhaps related to the orbital period; and
\citet{now99} reported a 117-day period in the X-ray light curve, although
this period appears to be evanescent \citep{wij02}.  
The distance to 4U~1957+115 is unknown and its mean X-ray luminosity
consequently uncertain; but for a distance of 7~kpc the mean 
unabsorbed luminosity 
is $\log(\textrm{L}_X) \approx 36.5$ (erg~sec$^{-1}$, 1.5-12 keV), 
while for a distance of
40 kpc it is $\log(\textrm{L}_X) \approx 38.1$ \citep{rus10}.

The X-ray spectrum of 4U~1957+115 has been measured numerous times.
It is unusually soft, placing 4U~1957+115 in the
region of the X-ray color-color diagram occupied by black
hole binaries, including black hole soft X-ray transients (SXTs)
in their outburst states, leading to the suspicion that 
4U~1957+115 harbors a black hole \citep{whi84, sch89, mcc03}.
If it does contain a black hole, it would be the only known black hole 
binary in the Galaxy that has not been seen to cycle between 
low and outburst states.
The nearly featureless spectrum can be successfully fitted 
with a variety of models that include some combination of 
three components:
a soft component from a single- or multi-temperature
black body, presumably coming from the accretion disk;  a hard
component that is either a power law, with or without a cutoff,
or a thermal Comptonization spectrum; and a broad emission 
feature near 6.5~keV from Fe K-line emission
\citep{yaq93,singh94,ric95,now99,wij02}.
The spectrum is softer when 4U~1957+115 is fainter, harder when 
it is brighter.

Most recently \citet{now08} fit high-quality Chandra, XMM-Newton,
and RXTE observations of 4U~1957+115 with {\tt diskbb} \citep{mit84} 
and {\tt kerrbb} models \citep[see][]{li05,dav06}
in the XSPEC package, fixing 
the inclination at $75^{\circ}$, and found that these black hole
models fit the observed X-ray spectrum well.   
The fits required high inner-disk temperatures,
$\textrm{kT}_{in} = 1.2 - 1.8\ \textrm{keV}$, and a low 
normalization factor,
$(R_{in}/D_{10})^2 \cos i = 5 - 25$, where $R_{in}$ is
the inner radius of the disk in km, $D_{10}$ is the distance
in units of 10 kpc, and $i$ is the orbital inclination.  
There are degeneracies among the derived values for the mass 
of the compact object, the mass accretion rate, and the 
distance, but the best fitting models ranged from a 
$3\ \textrm{M}_\odot$ black hole
with a spin of $a^{*}>0.82$ at a distance of 10~kpc to a 
$16\ \textrm{M}_\odot$ black hole with $a^{*}\approx 1$
at a distance of 22~kpc.
Nevertheless, the question remains: Is the compact object in 
4U~1957+115 really a black hole?  
The evidence for a black hole is hardly conclusive.
A soft spectrum does not necessarily require a 
black hole -- many SXTs contain neutron stars \citep{tan96},
and \citet{now08} did not explore models for disks around 
neutron stars.
Furthermore, 
\citet{yaq93} argued that the high values of $\textrm{kT}_{in}$
and low values of $R^2_{in}\cos i$ deduced from fits to the
X-ray spectral energy distribution are more typical
of neutron star systems than black hole systems.

In this paper we present new optical photometry of V1408~Aql.  
We refine the orbital period and derive the mean orbital
light curve.
There are no eclipses in the light curve.
We achieve good fits to the orbital light curve with a model similar to
that proposed by \citet{tho87} in which the
orbital modulation is due to the varying aspect of the irradiated face
of the secondary star.  
We believe this model is more physically realistic than models
that invoke a tall disk rim to produce the orbital modulation, but 
it increases the range of possible orbital inclinations and
mass ratios.

\section{Observations and Data Reduction}

\begin{figure}
\center 
\includegraphics[scale=0.65,angle=0]{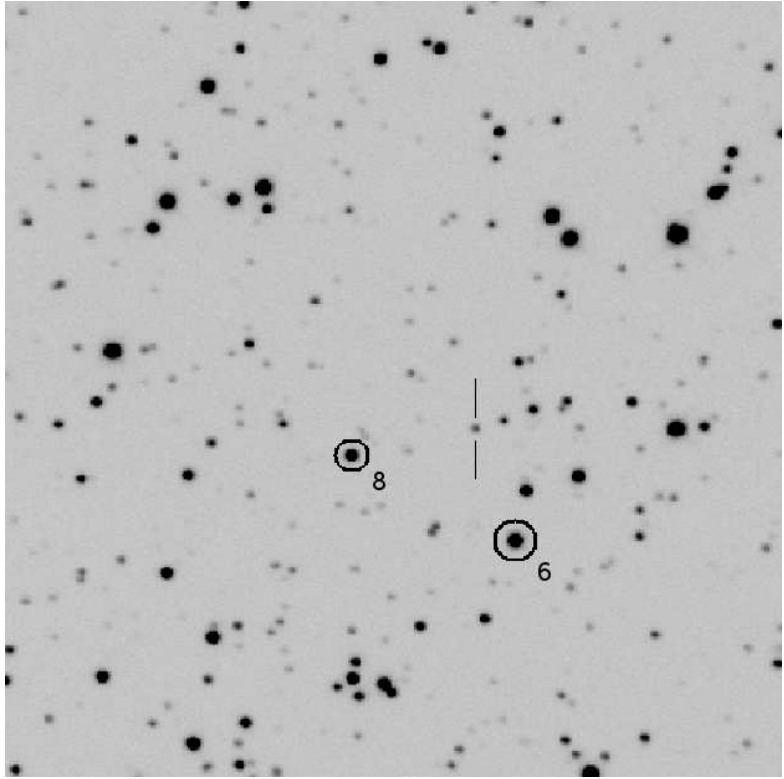}
\caption{A $BVR$ image of V1408~Aql and its surrounding field.  
V1408~Aql is identified with vertical hash marks and the 
two comparison stars are circled.  
The image dimensions are $2\farcm 8 \times 2 \farcm 8$ with north up 
and east to the left.}
\label{fc}
\end{figure}

We observed V1408~Aql from 2008 June 2 to June 7 UTC and again 
from August 2 to August 7 with the Argos prime-focus 
high-speed CCD photometer on the 2.1-m Otto Struve 
telescope at McDonald Observatory \citep{nat04}.  
We did not obtain data on June 6 nor on August 3 and 4 because
of clouds. 
The observations were made through a filter equivalent to 
a combined $BVR$ filter transmitting the
wavelength range 4130--7385~\AA\, and with an exposure time 
of 10 seconds for all frames. 
The data were reduced using a combination of
IRAF routines and an IDL pipeline (R. Hynes,
private communication).  
Data reduction for the Argos photometer is
discussed in detail in \citet{hyn04}. 
We measured relative fluxes by
dividing the flux from V1408~Aql by the total flux from two nearby
comparison stars in the frame.  
Figure \ref{fc} shows a stacked image of three $BVR$ images of 
the field of V1408~Aql for a 30-second effective exposure. 
V1408~Aql is indicated with vertical
hash marks and the two comparison stars are circled.  
These are the comparison stars numbered 6 and 8 in \citet{dox77,mar78}.

\section{The Optical Light Curve and the Orbital Period}

To search for orbital variations we folded the light curve on a 
linear ephemeris with a period of 9.331~h (see below).
Figure \ref{lc} shows the phase-folded data from each night binned 
into 350 equal-width phase bins.
The error bars shown in the figure are based on photon counting 
noise and do not
include other noise sources, such as accretion disk flickering.  
Observations with errors greater than
$\pm0.001$ in relative flux ($\sim$2.5\%) due to poor seeing or 
clouds have been discarded.
The light curve has a pronounced, roughly sinusoidal
orbital modulation.  
There are also flickering variations with an amplitude of 
$\sim 5\%$ of the median flux over an interval of $\sim0.15$
in phase or $\sim1.4$ hours.
There are no obvious eclipses. 

The mean flux from V1408~Aql changes on time scales of days
to months.
The system was fainter in June, brighter during the first two nights
in August, and fading on August 6 and 7.   
The top panel of Figure~\ref{lc}
shows the fainter light curve, the middle panel shows the 
brighter light curve, and the bottom panel shows the transition 
light curves.  
The color and symbol combinations indicate the different nights of
observations.   
The dashed line is a sine curve fitted to the June data 
(top panel) and plotted in the middle and bottom panels 
for comparison.
The brighter state has $\sim 30$\% more flux than the 
fainter state but
the peak-to-valley amplitude of the sinusoidal variations is 
nearly the same fraction of the median flux in both states.

The orbital light curve of V1408~Aql in 2008 was nearly 
identical to the light curve measured by \citet{tho87} in 1985.
In both years the variations of the mean light curve were 
nearly sinusoidal with, as we will show in section~5.2, peak-to-peak
amplitudes near 23\%.
While it is possible that the mean orbital 
light curve was truly different in 1996 when observed by
\citet{hak99}, they constructed the 1996 light curve
from data that were obtained on just two nights and
covered slightly less than one full orbit with no phase overlap.
It is more likely that the light curve was distorted by
flickering and night-to-night variations of the mean flux, 
and is not a good representation of the mean orbital light curve.
Since the variations of color with orbital phase are also affected
by these problems, it is premature to conclude that 
V1408~Aql becomes bluer at fainter orbital phases.
On the other hand, Figures~2 and 3 in \citet{hak99} 
do show that any color changes are small and, in the BVR passbands,
consistent with zero.

\citet{rus10} did not detect the orbital modulation at all in
photometry of V1408~Aql obtained between 2006 and 2009, 
and concluded that the system had changed.
Their sparse data set (eg, 35 data points in $V$ spread
over 3.1 years) is not, however, well suited for detecting an
orbital modulation that has been diluted by flickering and
night-to-night variations.
Also, their densest data set, from 2008, by chance covers 
only half the orbit with consequent loss of sensitivity to 
orbital modulations.
Our data, taken within a few weeks of theirs in 2008, shows
that the orbital modulation was clearly present.
The \citet{rus10} data does show that the night-to-night 
variations can have a total range of at least 0.8~mag, 
somewhat higher than the 0.6~mag we observed.

To refine the orbital period we first separately
calculated the medians of the fainter state and brighter state 
light curves and then subtracted the medians to place the 
faint and bright light curves at zero median flux.  
We then performed a Lomb-Scargle (L-S) period search, 
excluding the transition light curves \citep{lom76,sca82}.
The L-S periodogram in Figure~\ref{lomb} shows a series of
peaks near the orbital period found by \citet{tho87}, with the 
two highest peaks at $9.331\pm0.003$~h and $9.393 \pm 0.003$~h.
All these peaks have vanishingly small false alarm probabilities
($\ll 10^{-10}$),
reflecting the large amplitude of the orbital variation, and
all are two-month aliases of each other introduced by 
the two-month separation of our observing runs.
Because Thorstensen's data were taken over a single 12-day
interval and are not subject to one- or two-month aliases, 
we have used his period to choose among the aliases in our data. 
His period, $9.329\pm0.011$ h, agrees with the
$9.331\pm0.003$~h peak in our L-S periodogram to within a standard
deviation and differs from the other peaks in Figure~\ref{lomb}
by six or more standard deviations.
The orbital ephemeris is, then,
\begin{equation}
T = 2454621.86(02) + 0.38877(13)E,
\label{ephem}
\end{equation}
where $T(E=0)$ is the time of maximum flux.
We searched for, but did not find, any periodic modulations 
at other frequencies in the optical data, in agreement 
with \citet{tho87}.  

\section{The X-ray Light Curve and X-ray Periodicity Search}

Figure \ref{xte} shows the one-day-averaged X-ray light curve 
of 4U~1957+115 from 
the RXTE All Sky Monitor (ASM)\footnote{Data provided by 
the ASM/RXTE teams at MIT and at the RXTE SOF and GOF at 
NASA's GSFC.  http://xte.mit.edu} during 2008 May to August.  
The times of our optical observations are also 
shown in the figure.  
The X-ray flux was fairly constant in June when V1408~Aql was 
faint at optical wavelengths.  
The X-ray flux was generally higher in late July and August, 
but on those dates in August when we observed V1408~Aql, the 
X-ray flux was not any higher than it was in June. 

We searched for periodicities in the X-ray light curve 
from 1996 January to 2008 October in dwell-by-dwell samples 
using both a least-squares sine curve periodogram and 
a L-S periodogram.  
The L-S periodograms in Figures \ref{xls1} and \ref{xsine} are 
fully-generalized L-S periodograms \citep{zec09}. 
This method is an improvement over the original L-S periodogram 
because it includes the weights of individual data points 
and allows for non-zero means. 
The periodograms have been normalized assuming the noise is white.  
Figures \ref{xls1} and \ref{xsine} also show least squares sine fit 
periodograms for comparison.  

Figure \ref{xls1} shows the periodograms of the X-ray light curve 
at periods near the orbital period.  
The open circle with error bars marks a frequency 
of 2.5726 day$^{-1}$, corresponding to the orbital period from 
\citet{tho87}; 
and the dashed line marks a frequency of 2.5722 day$^{-1}$, 
corresponding to the 9.331-hour orbital period from our data.
The horizontal dashed line is drawn at the L-S power level 
for which the false alarm probability is 99\%.
The dot at 2.5758 day$^{-1}$ marks the ``marginal detection'' 
at $P=9.3175$ hours from \citet{lev06}.
There are no significant peaks in the periodogram at this 
or any other frequency between 2.54 and 2.6 day$^{-1}$.
Thus, there is no evidence for any modulation of the X-ray 
flux at periods near the orbital period. 
Figure \ref{xsine} shows the low frequency periodograms. 
The extra power at frequencies less than $\sim0.01$~d$^{-1}$
is real in the sense that the probability it could be produced
by white noise is $\ll 10^{-8}$.
None of the individual low-frequency peaks correspond to any 
of the long periods
mentioned by \citet{now99} and \citet{wij02}, including those
at 58.5~d, 117~d, and 234~d.
We therefore attribute this excess power to 
non-periodic or perhaps transient periodic variations of the 
X-ray flux.

\section{A Model for the Orbital Light Curve}

\subsection{Preliminary Considerations}
We modeled the orbital light curve of V1408~Aql using our 
light curve synthesis program\footnote{A full description of 
the program is available 
at \texttt{http://pisces.as.utexas.edu/robinson/XRbinary.pdf}.} 
for X-ray binaries \citep{bay10}. 
The program assumes that the orbital eccentricity is zero, 
the primary star is a point source surrounded by an accretion 
disk, and the secondary star fills its Roche Lobe. 
The accretion disk can have a complicated geometry.
It can be non-circular, non-axisymmetric, and
spotted, and can have structures that extend vertically out of the plane
of the disk, such as an interior torus and a tall disk rim.
The program uses stellar atmospheres from \citet{kur96} for the spectrum 
of the secondary star and blackbody spectra for all other structures.  
It allows for heating by irradiation but not for second-order 
heating by the irradiated surfaces. 
Light curves can be calculated for Johnson/Cousins filters or 
for a square bandpass over a specified wavelength range.   
Because we observed with a broadband $BVR$ filter, we used the 
square bandpass option with wavelength range 4130--7385~\AA.

The X-ray spectrum of 4U~1957+115 is remarkably simple.
\citet{now08} suggest that it ``...may be the cleanest disk
spectrum with which to study modern disk models.'' 
As noted in the introduction,
the continuum can be modeled with just a multi-temperature
black body disk, perhaps with a weak optically-thin
comptonizing corona.
With the possible exception of the Fe K-line, there are no 
strong emission lines.
The spectrum does have absorption lines, but \citet{now08} 
attribute the lines to absorption in hot-phase interstellar
material in the line of sight to  4U~1957+115.
The X-ray light curve shows no periodicities 
that might be attributable to disk warping, super-humps, 
precession, or other disk asymmetries.
Thus, neither the X-ray spectrum nor the X-ray light curve
require models with complex disk structures. 
We are also loathe to invoke a tall variable-height rim at 
the outer edge of the disk to modulate the optical 
orbital light curve.
As we have noted previously \citep{bay10}, tall rims are 
highly unphysical, requiring either supersonic 
turbulence or temperatures much higher than expected near
the edge of the disk.

We have, therefore, adopted a thin, axisymmetric disk in our 
model.
Since there are no eclipses, the only effect of
the disk is to add constant flux to the light curve and 
to irradiate the secondary star; and
the only source of orbital modulation is the varying
aspect of the irradiated secondary star.
This simple model has just five meaningful parameters: the mass ratio,
the orbital inclination, the zero point in orbital phase, 
the X-ray flux from the central source, which heats 
the secondary star, and an additional constant optical flux 
from the disk, which also heats the secondary star.  

\subsection{Fits to the Light Curve}
A preliminary analysis showed that
the model yields synthetic light curves that fit
the observed orbital light curves equally well for a 
wide range of parameters.
We therefore abandoned any attempt to define
``best'' values for the mass ratio and orbital inclination and,
instead, explored the range of parameters
that yield acceptable fits.
We considered three mass ratios, $q=M_2/M_{\rm X}=0.3$, 0.1, and 0.025, 
roughly corresponding to a compact object that is a 
neutron star, a low mass black hole, and a  high mass black hole.  
For each mass ratio we fit synthetic light curves for models 
with inclinations 
of $10^{\circ}$, $20^{\circ}$, $50^{\circ}$, $65^{\circ}$, 
and $70^{\circ}$, plus some for models 
at $5^{\circ}$, $75^{\circ}$, and $80^{\circ}$. 
For concreteness, we set the mass of the secondary star to 
$M_2 = 0.4\ M_\odot$ and its temperature to 3500~K.
Our results are nearly independent of $M_2$
since the geometry of the system depends only on $q$ and 
the scale of the system depends only on $M_2^{1/3}$.
The secondary's low temperature prevents its un-irradiated 
surfaces from adding significant flux to the light curve. 

We quantified the quality of the fits with $\chi^2$ and used 
the simplex algorithm \citep{nel65} in IDL to find the 
values of the remaining three parameters (zero phase,
X-ray flux, optical flux) that minimized $\chi^2$.
Because of the flickering, light curves from individual nights 
often depart from the mean orbital light curve by much more than the 
standard deviation of the measurements, which greatly increases 
$\chi^2$.
Since the flickering in effect adds non-white noise to the data 
and the light curves are
sampled with a high cadence, the departures are strongly correlated, 
thwarting our attempts to artificially increase the variances of the 
individual data points to account for the flickering.
As a result, the absolute value of $\chi^2$ loses its meaning.
Nevertheless, the relative value of $\chi^2$ retains its 
utility for measuring the relative quality of fits.

Figures \ref{nslc} and \ref{hmlc} show the observed
light curves overlayed by fitted synthetic 
light curves.
The figures show fits to the light curves during
both the fainter and brighter states,
for two inclinations near the limits of the range we
explored, $i=20^{\circ}$ and $i=65^{\circ}$, and for
the largest and smallest mass ratios we explored, $q=0.3$ 
(Figure \ref{nslc}) and $q=0.025$ (Figure \ref{hmlc}).
The peak-to-peak amplitude of the fitted synthetic light 
curve is 23\% of the mean flux when V1408~Aql is in its fainter 
state and 24\% of the mean flux when it is in its brighter state.
For comparison, \citet{tho87} measured an amplitude of 
0.116~mag or a peak-to-peak amplitude of 22.6\%.
The independence of the fractional amplitude from the mean flux is an
important constraint.
We interpret the constraint to mean that (1) the
optical flux from both the disk and the secondary star
is dominated by reprocessed X-ray flux and (2) the flux-weighted 
temperatures of the irradiated surfaces
are high enough that the emitted optical flux is on the Rayleigh-Jeans
tail of the blackbody distribution.
Changes in the X-ray flux then produce proportional 
changes in the optical flux from the disk and from the 
heated face of the secondary star, leaving their relative 
contributions to the optical flux unchanged.
The blue colors of V1408~Aql 
($B-V \approx 0.13$, $U-B \approx -0.70$, $E(B-V) \approx 0.3$ 
\citep{mar78,hak99}) 
are at least roughly consistent with this interpretation.

We achieved equally good fits to the orbital light curves for all
three mass ratios.
The dominant effect of increasing the mass ratio is to increase
the relative size of the secondary star; but
the effect of the increased size can be offset by reducing
the amount of irradiative heating.
As a result, the observed light curves do not by themselves
yield useful limits on the mass ratio.

We have two ways to place an upper limit on the orbital inclination.
The first is from the lack of eclipses in the light curve.
Let us assume that the accretion disk extends to its tidal 
truncation radius, $R_d = 0.9 R_{L1}$, where $R_{L1}$ is
the radius of a sphere with the same volume as the Roche
lobe around the compact object.
Eclipses of the disk would be detected in the
mean light curve if they were more than about 2\% deep.
This occurs at $i \gtrsim 65^{\circ}$ for $q = 0.3$,
at $i \gtrsim 70^{\circ}$ for $q = 0.1$, and at
$i \gtrsim 75^{\circ}$ for $q = 0.025$.
The limiting inclinations are higher if $R_d < 0.9 R_{L1}$.
The second way to place an upper limit on the inclination
is from a flattening of the minimum of the
synthetic light curves at high orbital inclinations.
Figure~\ref{70ls} shows synthetic light curves for
$q=0.3$ (top, red line), $q=0.1$ (middle, green line), 
and $q=0.025$ (bottom, blue line), each for an inclination 
of $70^{\circ}$.
The synthetic light curves all flatten at minimum light. 
This flattening is unrelated to eclipses but instead 
occurs because less than half the secondary star is irradiated,
because the edges of the irradiated region are heated to
lower temperatures, and because the secondary is not spherical. 
The lack of flat minima in the observed 
optical light curves places an upper limit of $70^{\circ}$
on the orbital inclination.
This second limit is independent of mass ratio and 
the size of the accretion disk, and is more reliable than
the limits set by the lack of disk eclipses.

The synthetic light 
curves for models with orbital inclinations of $20^{\circ}$ and
$65^{\circ}$ both fit the observed light curves well.
The fit for $20^{\circ}$ has a slightly smaller $\chi^2$ but the 
difference is not statistically significant.  
The dominant effect of decreasing the orbital inclination is
to decrease the amplitude of the orbital modulation caused by
the changing aspect of the secondary star, but
the decrease of the relative amplitude can be offset 
by decreasing the fraction of the flux contributed by
the accretion disk.
This is shown in Table~1, which gives the fraction of the 
optical flux, $\alpha$, that comes from the accretion disk as 
a function of orbital inclination;
the fraction is also shown as the dashed lines in
Figures \ref{nslc} and \ref{hmlc}.
The value of $\alpha$ is nearly independent of the mass ratio and
the luminosity state, but decreases with orbital inclination,
dropping slowly from $\sim 88\%$ at 
$i = 65^{\circ}$ to $\sim 67\%$ at $i = 20^{\circ}$, and then
dropping rapidly to zero at $\sim 5^{\circ}$.

The flickering in the light curve disfavors inclinations
less than $20^{\circ}$.
The low values of $\alpha$ at $i < 20^{\circ}$ would require
an improbably high fraction of the disk flux to be modulated
to produce the observed amplitude of the flickering.

\subsection{Constraints on ${\mathbf q}$ and ${\mathbf i}$ from Color and Temperature}

The results from the previous section show that neither
the mass ratio nor the orbital inclination are 
strongly constrained by the shape of the light curve.
They are, however, constrained by color and temperature,
albeit rather weakly and with some uncertainty.
We take as our starting point 
that the optical flux from V1408~Aql is dominated by regions 
of the secondary star and disk that are on the Rayleigh-Jeans
tail of the black body distribution.
Let $T_2$ and $T_d$ be the flux weighted temperatures of
the irradiated areas of the secondary star and disk so that
$T_2/T_d$ is the ratio of their emitted optical fluxes.
We have two ways to calculate $T_2/T_d$.
The first is from the amplitude of the orbital modulation,
which is nearly independent of the luminosity and linear scale 
of the model;
and the second is from the irradiative heating of the disk
and secondary star, which depends strongly on the luminosity and scale
of the model.
The two calculations must agree for the
model to be internally consistent.

From the amplitude of the orbital modulation,
the ratio of the optical energy coming from the secondary to
the optical energy from the disk is
\begin{equation}
  \left( {{T_2} \over {T_d}} \right)
  \left( {{A_2 / 2} \over{A_d \cos i}} \right)
          \ \approx \ {{1-\alpha} \over {\alpha}},
  \label{fluxratio-eq}
\end{equation}
where $A_2 / 2$ and $A_d \cos i$ are the areas of the secondary
star and disk that contribute significantly to the optical flux,
the factor of 2 because only half the secondary is irradiated
and the factor of $\cos i$ because the disk is foreshortened.
If the disk has expanded to its tidal truncation 
radius, then $R_d = 0.9 R_{L1}$, and
equation~\ref{fluxratio-eq} can be rewritten as
\begin{equation}
  {{T_2} \over {T_d}} \ = \ {{1-\alpha} \over {\alpha}}
                 \left[
                 {{0.81 (R_{L1}/a)^2 \cos i} \over {0.5 (R_{L2}/a)^2}}
                 \right] ,
\end{equation}
where $a$ is the semi-major axis, and $R_{L1}/a$ and $R_{L2}/a$
are functions of $q$ and can be calculated from, for example, 
Eggleton's approximation \citep{egg83}.
Table~1 gives the values of $T_2/T_d$ implied by the
fits to the light curves.
Note that $T_2/T_d$ must be larger for smaller mass ratios because
the relative size of the secondary star is smaller.

The temperature of the irradiated secondary depends primarily on 
$q$, the scale of the system (set by $M_2$), the X-ray
luminosity, and the fraction of the incident radiation that
is locally thermalized and re-emitted.
Table~2 gives typical values for $T_2$ for
X-ray luminosities of $10^{37}$ and $10^{38}$~erg s$^{-1}$
for each of the three mass ratios, where we have assumed that 
the X-ray energy is produced near the compact object and
that all the X-ray energy incident on the secondary star
is thermalized and re-emitted.
The temperature of an irradiated disk is extremely sensitive
to disk geometry and the location of the source of the 
irradiating X-rays (see, eg, \citet{hyn05}).
As a result, our models give no reliable constraint on $T_d$.
We will simply take that $T_d$ must be greater than $\sim 10^4$~K 
for its optical flux to be on the Rayleigh-Jeans tail of the 
blackbody distribution.
It is possible, of course, that $T_d$ is substantially 
greater than $10^4$~K.

We can now calculate $T_2/T_d$ the second way, 
using $T_2$ from Table~2 and taking $T_d > 10^4$~K
so that the resulting $T_2/T_d$ is an upper limit.
For $q = 0.3$, the upper limits are
$T_2/T_d < 5$ for $L_X = 10^{38}$~erg~s$^{-1}$ and 
$T_2/T_d < 2.8$ for $10^{37}$~erg~s$^{-1}$
These values are consistent with the
values in Table~1 for any $i \gtrsim 20^{\circ}$ and
yield no additional useful constraint on the inclination.
For $q = 0.1$ the upper limits are 4.0 and 2.2 and 
are mildly constrained, needing 
$i \gtrsim 30^{\circ}$ for $10^{38}$~erg s$^{-1}$ and 
$i \gtrsim 40^{\circ}$ for $10^{37}$~erg s$^{-1}$.
At $q = 0.025$ the upper limits are 3.1 and 1.7 and become
more strongly constraining.
Only the model with $L_X = 10^{38}$~erg s$^{-1}$ is 
internally consistent and even it requires $i \gtrsim 60^{\circ}$.

These constraints are not, however, as certain as one might wish.
The values of $T_2/T_d$ given in Table~1 are directly
proportional to the adopted area of the accretion disk.
If the radius of the accretion disk is as small as the 
circularization radius, $T_2/T_d$ is reduced by a factor 
of 4 or more, and the inclination becomes essentially 
unconstrained even for $q = 0.025$.
On the other hand, if $T_d$ is greater than $10^4$~K, the 
constraints on the orbital inclination become more severe.
Thus the most that can be concluded is that higher mass 
ratios are mildly favored and the models are internally 
consistent over a wide range of inclinations and mass ratios. 

Finally, it is possible that further observations may show that
V1408~Aql does, indeed, becomes bluer at fainter orbital 
phases as claimed by \citet{hak99}.
Our model can reproduce such color variations -- all that
is required is that the heated face of the secondary star 
have a temperature less than the temperature of the accretion disk.
This would strongly constrain the permitted mass ratios
and inclinations, eliminating
the $q = 0.025$ models altogether and restricting the
$q = 0.1$ models to inclinations greater than $\sim 65^{\circ}$
and the $q = 0.3$ models to inclinations greater than
$\sim 40^{\circ}$.

\section{Summary and Discussion}

We obtained high-speed photometry of V1408~Aql on nine nights 
in June and August of 2008.
The optical light curve flickered by at least 5\% on time scales of
minutes to hours and the orbit-averaged flux varied by 30\% on time 
scales of a few days. 
The mean orbital light curve was roughly sinusoidal
with a peak-to-peak amplitude near 23\% and was not measurably
different from the light curve observed by \citet{tho87} in
1985.
We attribute the lack of an orbital modulation reported 
by \citet{rus10} and the different shape of the
orbital light curve found by \citet{hak99} to the
sparseness of their data sets and to the distortions of the light curve
induced by flickering and night-to-night variations.
The refined orbital period is $0.38877\pm0.00013$~d 
(equation~\ref{ephem}).
There are no significant periodicities in
the RXTE ASM X-ray light curve at or near the
orbital period.

The orbital light curve is consistent with a model 
in which the accretion disk around the compact star
is thin, axisymmetric, and uneclipsed.
The secondary star is irradiated and heated by X-rays from the compact
star and the accretion disk.
The orbital modulation is caused entirely by the varying 
aspect of the heated face of the secondary star with orbital phase.  
This simple model yields consistent fits for all mean 
light levels and it avoids an unphysically tall disk rim.   

At inclinations greater than $70^{\circ}$ the synthetic 
orbital light curves flatten at minimum light.  
The lack of a flat minimum in the observed 
light curve places an upper limit of $\sim 70^{\circ}$ 
on the orbital inclination. 
We cannot place an unimpeachable lower limit on the
orbital inclination but
argue that both the amplitude of the flickering and
the lack of color variations over the orbit imply that
the orbital inclination is greater than $20^{\circ}$.   
Thus, the orbital inclination lies between
$20^{\circ}$ and $70^{\circ}$.
We achieved good fits to the light curves with mass ratios
from 0.025 to 0.3.
If the radius of the accretion disk is close to the tidal
truncation radius, the smaller mass ratios, which correspond
to systems in which the primary is a black hole, require 
X-ray luminosities nearer $10^{38}$~erg~s$^{-1}$ and 
orbital inclinations in the upper half of the permitted
range.

Figure 8 in \citet{rus10} plots optical luminosity against 
X-ray luminosity for V1408~Aql and other LMXBs.  
The distance to V1408~Aql is unknown, but \citet{rus10} 
adopt three distances for comparison.  
If the distance is 7~kpc, the X-ray luminosity 
is $\sim 10^{36.5}$~erg~s$^{-1}$, and at  
at 20~kpc, it is $\sim 10^{37.5}$~erg~s$^{-1}$,
placing V1408~Aql in a regions of the figure occupied by neutron
star binaries but not black hole binaries.
At 40~kpc, the X-ray luminosity is
$\sim 10^{38.5}$~erg~s$^{-1}$ and V1408~Aql falls in
a region occupied by some soft-state black hole binaries
and the neutron star Z-sources.  
All these possibilities remain viable with our new model. 

If, however, V1408~Aql contains a black hole it would be unusual
in two ways.
First, as mentioned earlier it would be the only known black hole 
binary in the Galaxy that has not been seen to cycle between 
low and outburst states.
Second, 
irradiative heating is more typical of neutron star X-ray binaries than
black hole binaries.
This is shown most clearly by SXTs during their outbursts.
Large amplitude modulations from irradiative heating have been
observed during outbursts of Aql~X-1 and XTE~2123-058 and possibly X1608-52,
all of which are classical SXTs containing neutron stars;
and have also been observed in the related neutron star X-ray transient
XTE~J2129+47 \citep{tho79,wel00,zur00,wac02}.
In contrast, orbital modulation from irradiative
heating of the secondary star is rarely observed in black hole 
SXTs and, if it is observed, its amplitude is low \citep{nei07}.
This is due at least partly to the lower mass ratios of
black hole binaries.
At lower mass ratios the secondary star is comparatively smaller.
As shown in section 5.3, the smaller secondary can produce a 
large-amplitude orbital modulation only if its irradiated face 
is heated to high temperatures.

Finally, the light curves of the black hole SXTs often show 
superhumps \citep{has01,zur06,kat09}.
Such superhumps are ubiquitous in mass transfer binaries with
disks whose radii cross the 3:1 orbital resonance, which requires a
mass ratio less than $\sim 0.25$ \citep{lub91,kat09}.
The failure of V1408~Aql to show superhumps in its light curve argues
that its mass ratio is larger than $\sim 0.25$ and the mass
of its primary star is less than $\sim 1.6\ M_\odot$.
All the known black holes in binary
stars have masses greater than about $4\ M_\odot$ \citep{cas06}.
While none of these arguments are conclusive, together they
weigh heavily in favor of a neutron star and against a black 
hole in V1408~Aql.

We thank R. I. Hynes for helpful discussions.


\newpage
\begin{thebibliography}{39}
\expandafter\ifx\csname natexlab\endcsname\relax\def\natexlab#1{#1}\fi

\bibitem[{{Bayless} {et~al.}(2010){Bayless}, {Robinson}, {Hynes}, {Ashcraft},
  \& {Cornell}}]{bay10}
{Bayless}, A.~J., {Robinson}, E.~L., {Hynes}, R.~I., {Ashcraft}, T.~A., \&
  {Cornell}, M.~E. 2010, \apj, 709, 251

\bibitem[{{Casares}(2006)}]{cas06}
{Casares}, J. 2006, in The Many Scales in the Universe: JENAM 2004 Astrophysics
  Reviews, ed. {J.~C.~Del Toro Iniesta, E.~J.~Alfaro, J.~G.~Gorgas,
  E.~Salvador-Sole, \& H.~Butcher}, 145--+

\bibitem[{{Davis} {et~al.}(2006){Davis}, {Done}, \& {Blaes}}]{dav06}
{Davis}, S.~W., {Done}, C., \& {Blaes}, O.~M. 2006, \apj, 647, 525

\bibitem[{{Doxsey} {et~al.}(1977){Doxsey}, {Bradt}, {Dower}, {Jernigan}, \&
  {Apparao}}]{dox77}
{Doxsey}, R.~E., {Bradt}, H.~V., {Dower}, R.~G., {Jernigan}, J.~G., \&
  {Apparao}, K.~M.~V. 1977, \nat, 269, 112

\bibitem[{{Eggleton}(1983)}]{egg83}
{Eggleton}, P.~P. 1983, \apj, 268, 368

\bibitem[{{Hakala} {et~al.}(1999){Hakala}, {Muhli}, \& {Dubus}}]{hak99}
{Hakala}, P.~J., {Muhli}, P., \& {Dubus}, G. 1999, \mnras, 306, 701

\bibitem[{{Haswell} {et~al.}(2001){Haswell}, {King}, {Murray}, \&
  {Charles}}]{has01}
{Haswell}, C.~A., {King}, A.~R., {Murray}, J.~R., \& {Charles}, P.~A. 2001,
  \mnras, 321, 475

\bibitem[{{Hynes}(2005)}]{hyn05}
{Hynes}, R.~I. 2005, \apj, 623, 1026

\bibitem[{{Hynes} {et~al.}(2004){Hynes}, {Robinson}, \& {Jeffery}}]{hyn04}
{Hynes}, R.~I., {Robinson}, E.~L., \& {Jeffery}, E. 2004, \apjl, 608, L101

\bibitem[{{Kato} {et~al.}(2009){Kato}, {Imada}, {Uemura}, {Nogami}, {Maehara},
  {Ishioka}, {Baba}, {Matsumoto}, {Iwamatsu}, {Kubota}, {Sugiyasu}, {Soejima},
  {Moritani}, {Ohshima}, {Ohashi}, {Tanaka}, {Sasada}, {Arai}, {Nakajima},
  {Kiyota}, {Tanabe}, {Imamura}, {Kunitomi}, {Kunihiro}, {Taguchi}, {Koizumi},
  {Yamada}, {Nishi}, {Kida}, {Tanaka}, {Ueoka}, {Yasui}, {Maruoka}, {Henden},
  {Oksanen}, {Moilanen}, {Tikkanen}, {Aho}, {Monard}, {Itoh}, {Dubovsky},
  {Kudzej}, {Dancikova}, {Vanmunster}, {Pietz}, {Bolt}, {Boyd}, {Nelson},
  {Krajci}, {Cook}, {Torii}, {Starkey}, {Shears}, {Jensen}, {Masi}, {Hynek},
  {Nov{\'a}}, {K}, {Koci{\'a}}, {N}, {Kr{\'a}}, {L}, {Ku{\v c}{\'a}},
  {Kov{\'a}}, {Kolasa}, {{\v S}tastn{\'y}}, {Staels}, {Miller}, {Sano}, {de
  Ponthi{\`e}re}, {Miyashita}, {Crawford}, {Brady}, {Santallo}, {Richards},
  {Martin}, {Buczynski}, {Richmond}, {Kern}, {Davis}, {Crabtree}, {Beaulieu},
  {Davis}, {Aggleton}, {Morelle}, {Pavlenko}, {Andreev}, {Baklanov},
  {Koppelman}, {Billings}, {Urbancok}, {{\"O}gmen}, {Heathcote}, {Gomez},
  {Voloshina}, {Retter}, {Mularczyk}, {Zoczewski}, {Olech}, {Kedzierski},
  {Pickard}, {Stockdale}, {Virtanen}, {Morikawa}, {Hambsch}, {Garradd},
  {Gualdoni}, {Geary}, {Omodaka}, {Sakai}, {Michel}, {C{\'a}rdenas}, {Gazeas},
  {Niarchos}, {Yushchenko}, {Mallia}, {Fiaschi}, {Good}, {Walker}, {James},
  {Douzu}, {Julian}, {Butterworth}, {Shugarov}, {Volkov}, {Chochol},
  {Katysheva}, {Rosenbush}, {Khramtsova}, {Kehusmaa}, {Reszelski}, {Bedient},
  {Liller}, {Pojmanski}, {Simonsen}, {Stubbings}, {Schmeer}, {Muyllaert},
  {Kinnunen}, {Poyner}, {Ripero}, \& {Kriebel}}]{kat09}
{Kato}, T., {Imada}, A., {Uemura}, M., {Nogami}, D., {Maehara}, H., {Ishioka},
  R., {Baba}, H., {Matsumoto}, K., {Iwamatsu}, H., {Kubota}, K., {Sugiyasu},
  K., {Soejima}, Y., {Moritani}, Y., {Ohshima}, T., {Ohashi}, H., {Tanaka}, J.,
  {Sasada}, M., {Arai}, A., {Nakajima}, K., {Kiyota}, S., {Tanabe}, K.,
  {Imamura}, K., {Kunitomi}, N., {Kunihiro}, K., {Taguchi}, H., {Koizumi}, M.,
  {Yamada}, N., {Nishi}, Y., {Kida}, M., {Tanaka}, S., {Ueoka}, R., {Yasui},
  H., {Maruoka}, K., {Henden}, A., {Oksanen}, A., {Moilanen}, M., {Tikkanen},
  P., {Aho}, M., {Monard}, B., {Itoh}, H., {Dubovsky}, P.~A., {Kudzej}, I.,
  {Dancikova}, R., {Vanmunster}, T., {Pietz}, J., {Bolt}, G., {Boyd}, D.,
  {Nelson}, P., {Krajci}, T., {Cook}, L.~M., {Torii}, K., {Starkey}, D.~R.,
  {Shears}, J., {Jensen}, L., {Masi}, G., {Hynek}, T., {Nov{\'a}}, {K}, R.,
  {Koci{\'a}}, {N}, R., {Kr{\'a}}, {L}, L., {Ku{\v c}{\'a}}, {Kov{\'a}}, H.,
  {Kolasa}, M., {{\v S}tastn{\'y}}, P., {Staels}, B., {Miller}, I., {Sano}, Y.,
  {de Ponthi{\`e}re}, P., {Miyashita}, A., {Crawford}, T., {Brady}, S.,
  {Santallo}, R., {Richards}, T., {Martin}, B., {Buczynski}, D., {Richmond},
  M., {Kern}, J., {Davis}, S., {Crabtree}, D., {Beaulieu}, K., {Davis}, T.,
  {Aggleton}, M., {Morelle}, E., {Pavlenko}, E.~P., {Andreev}, M., {Baklanov},
  A., {Koppelman}, M.~D., {Billings}, G., {Urbancok}, L., {{\"O}gmen}, Y.,
  {Heathcote}, B., {Gomez}, T.~L., {Voloshina}, I., {Retter}, A., {Mularczyk},
  K., {Zoczewski}, K., {Olech}, A., {Kedzierski}, P., {Pickard}, R.~D.,
  {Stockdale}, C., {Virtanen}, J., {Morikawa}, K., {Hambsch}, F., {Garradd},
  G., {Gualdoni}, C., {Geary}, K., {Omodaka}, T., {Sakai}, N., {Michel}, R.,
  {C{\'a}rdenas}, A.~A., {Gazeas}, K.~D., {Niarchos}, P.~G., {Yushchenko},
  A.~V., {Mallia}, F., {Fiaschi}, M., {Good}, G.~A., {Walker}, S., {James}, N.,
  {Douzu}, K., {Julian}, II, W.~M., {Butterworth}, N.~D., {Shugarov}, S.~Y.,
  {Volkov}, I., {Chochol}, D., {Katysheva}, N., {Rosenbush}, A.~E.,
  {Khramtsova}, M., {Kehusmaa}, P., {Reszelski}, M., {Bedient}, J., {Liller},
  W., {Pojmanski}, G., {Simonsen}, M., {Stubbings}, R., {Schmeer}, P.,
  {Muyllaert}, E., {Kinnunen}, T., {Poyner}, G., {Ripero}, J., \& {Kriebel}, W.
  2009, \pasj, 61, 395

\bibitem[{{Kurucz}(1996)}]{kur96}
{Kurucz}, R.~L. 1996, in Astronomical Society of the Pacific Conference Series,
  Vol. 108, M.A.S.S., Model Atmospheres and Spectrum Synthesis, ed.
  {S.~J.~Adelman, F.~Kupka, \& W.~W.~Weiss}, 2--+

\bibitem[{{Levine} \& {Corbet}(2006)}]{lev06}
{Levine}, A.~M. \& {Corbet}, R. 2006, The Astronomer's Telegram, 940, 1

\bibitem[{{Li} {et~al.}(2005){Li}, {Zimmerman}, {Narayan}, \&
  {McClintock}}]{li05}
{Li}, L., {Zimmerman}, E.~R., {Narayan}, R., \& {McClintock}, J.~E. 2005,
  \apjs, 157, 335

\bibitem[{{Lomb}(1976)}]{lom76}
{Lomb}, N.~R. 1976, \apss, 39, 447

\bibitem[{{Lubow}(1991)}]{lub91}
{Lubow}, S.~H. 1991, \apj, 381, 259

\bibitem[{{Margon} {et~al.}(1978){Margon}, {Thorstensen}, \& {Bowyer}}]{mar78}
{Margon}, B., {Thorstensen}, J.~R., \& {Bowyer}, S. 1978, \apj, 221, 907

\bibitem[{{McClintock} \& {Remillard}(2003)}]{mcc03}
{McClintock}, J.~E. \& {Remillard}, R.~A. 2003, Compact Stellar X-ray Sources,
  eds. W.H.G. Lewin and M. van der Klis (Cambridge University Press), 157

\bibitem[{{Mitsuda} {et~al.}(1984){Mitsuda}, {Inoue}, {Koyama}, {Makishima},
  {Matsuoka}, {Ogawara}, {Suzuki}, {Tanaka}, {Shibazaki}, \& {Hirano}}]{mit84}
{Mitsuda}, K., {Inoue}, H., {Koyama}, K., {Makishima}, K., {Matsuoka}, M.,
  {Ogawara}, Y., {Suzuki}, K., {Tanaka}, Y., {Shibazaki}, N., \& {Hirano}, T.
  1984, \pasj, 36, 741

\bibitem[{{Nather} \& {Mukadam}(2004)}]{nat04}
{Nather}, R.~E. \& {Mukadam}, A.~S. 2004, \apj, 605, 846

\bibitem[{{Neil} {et~al.}(2007){Neil}, {Bailyn}, \& {Cobb}}]{nei07}
{Neil}, E.~T., {Bailyn}, C.~D., \& {Cobb}, B.~E. 2007, \apj, 657, 409

\bibitem[{{Nelder} \& {Mead}(1965)}]{nel65}
{Nelder}, J.~A. \& {Mead}, R. 1965, Computer Journal, 7, 308

\bibitem[{{Nowak} {et~al.}(2008){Nowak}, {Juett}, {Homan}, {Yao}, {Wilms},
  {Schulz}, \& {Canizares}}]{now08}
{Nowak}, M.~A., {Juett}, A., {Homan}, J., {Yao}, Y., {Wilms}, J., {Schulz},
  N.~S., \& {Canizares}, C.~R. 2008, \apj, 689, 1199

\bibitem[{{Nowak} \& {Wilms}(1999)}]{now99}
{Nowak}, M.~A. \& {Wilms}, J. 1999, \apj, 522, 476

\bibitem[{{Ricci} {et~al.}(1995){Ricci}, {Israel}, \& {Stella}}]{ric95}
{Ricci}, D., {Israel}, G.~L., \& {Stella}, L. 1995, \aap, 299, 731

\bibitem[{{Russell} {et~al.}(2010){Russell}, {Lewis}, {Roche}, {Clark},
  {Breedt}, \& {Fender}}]{rus10}
{Russell}, D.~M., {Lewis}, F., {Roche}, P., {Clark}, J.~S., {Breedt}, E., \&
  {Fender}, R.~P. 2010, \mnras, 44

\bibitem[{{Scargle}(1982)}]{sca82}
{Scargle}, J.~D. 1982, \apj, 263, 835

\bibitem[{{Schulz} {et~al.}(1989){Schulz}, {Hasinger}, \& {Truemper}}]{sch89}
{Schulz}, N.~S., {Hasinger}, G., \& {Truemper}, J. 1989, \aap, 225, 48

\bibitem[{{Singh} {et~al.}(1994){Singh}, {Apparao}, \& {Kraft}}]{singh94}
{Singh}, K.~P., {Apparao}, K.~M.~V., \& {Kraft}, R.~P. 1994, \apj, 421, 753

\bibitem[{{Tanaka} \& {Shibazaki}(1996)}]{tan96}
{Tanaka}, Y. \& {Shibazaki}, N. 1996, \araa, 34, 607

\bibitem[{{Thorstensen} {et~al.}(1979){Thorstensen}, {Charles}, {Bowyer},
  {Briel}, {Doxsey}, {Griffiths}, \& {Schwartz}}]{tho79}
{Thorstensen}, J., {Charles}, P., {Bowyer}, S., {Briel}, U.~G., {Doxsey},
  R.~E., {Griffiths}, R.~E., \& {Schwartz}, D.~A. 1979, \apjl, 233, L57

\bibitem[{{Thorstensen}(1987)}]{tho87}
{Thorstensen}, J.~R. 1987, \apj, 312, 739

\bibitem[{{Wachter} {et~al.}(2002){Wachter}, {Hoard}, {Bailyn}, {Corbel}, \&
  {Kaaret}}]{wac02}
{Wachter}, S., {Hoard}, D.~W., {Bailyn}, C.~D., {Corbel}, S., \& {Kaaret}, P.
  2002, \apj, 568, 901

\bibitem[{{Welsh} {et~al.}(2000){Welsh}, {Robinson}, \& {Young}}]{wel00}
{Welsh}, W.~F., {Robinson}, E.~L., \& {Young}, P. 2000, \aj, 120, 943

\bibitem[{{White} \& {Marshall}(1984)}]{whi84}
{White}, N.~E. \& {Marshall}, F.~E. 1984, \apj, 281, 354

\bibitem[{{Wijnands} {et~al.}(2002){Wijnands}, {Miller}, \& {van der
  Klis}}]{wij02}
{Wijnands}, R., {Miller}, J.~M., \& {van der Klis}, M. 2002, \mnras, 331, 60

\bibitem[{{Yaqoob} {et~al.}(1993){Yaqoob}, {Ebisawa}, \& {Mitsuda}}]{yaq93}
{Yaqoob}, T., {Ebisawa}, K., \& {Mitsuda}, K. 1993, \mnras, 264, 411

\bibitem[{{Zechmeister} \& {K{\"u}rster}(2009)}]{zec09}
{Zechmeister}, M. \& {K{\"u}rster}, M. 2009, \aap, 496, 577

\bibitem[{{Zurita} {et~al.}(2000){Zurita}, {Casares}, {Shahbaz}, {Charles},
  {Hynes}, {Shugarov}, {Goransky}, {Pavlenko}, \& {Kuznetsova}}]{zur00}
{Zurita}, C., {Casares}, J., {Shahbaz}, T., {Charles}, P.~A., {Hynes}, R.~I.,
  {Shugarov}, S., {Goransky}, V., {Pavlenko}, E.~P., \& {Kuznetsova}, Y. 2000,
  \mnras, 316, 137

\bibitem[{{Zurita} {et~al.}(2006){Zurita}, {Torres}, {Steeghs},
  {Rodr{\'{\i}}guez-Gil}, {Mu{\~n}oz-Darias}, {Casares}, {Shahbaz},
  {Mart{\'{\i}}nez-Pais}, {Zhao}, {Garcia}, {Piccioni}, {Bartolini},
  {Guarnieri}, {Bloom}, {Blake}, {Falco}, {Szentgyorgyi}, \&
  {Skrutskie}}]{zur06}
{Zurita}, C., {Torres}, M.~A.~P., {Steeghs}, D., {Rodr{\'{\i}}guez-Gil}, P.,
  {Mu{\~n}oz-Darias}, T., {Casares}, J., {Shahbaz}, T., {Mart{\'{\i}}nez-Pais},
  I.~G., {Zhao}, P., {Garcia}, M.~R., {Piccioni}, A., {Bartolini}, C.,
  {Guarnieri}, A., {Bloom}, J.~S., {Blake}, C.~H., {Falco}, E.~E.,
  {Szentgyorgyi}, A., \& {Skrutskie}, M. 2006, \apj, 644, 432

\end{thebibliography}

\newpage

\begin{table}
\begin{center}
{Table 1: Properties of the Fitted Models} \\ [9pt]
\begin{tabular}{ccccccc}
\hline\hline
\\ [-9pt]
            &\multicolumn{2}{c}{Fraction of Flux from Disk} && \multicolumn{3}{c}{Inferred$^\dag$ $T_2/T_d$} \\ [3pt]
                                         \cline{2-3}  \cline{5-7} \\ [-9pt]
Inclination & Faint State & Bright State && $q=0.3$ & $q = 0.1$ & $q = 0.025$\\ [6pt]
\hline
\\ [-9pt]
$10^\circ$  &    0.337    &   0.334      &&   9.49   &    24.9   &    78.0   \\
$20^\circ$  &    0.683    &   0.666      &&   2.28   &    5.97   &    18.7   \\
$30^\circ$  &    0.784    &   0.759      &&   1.33   &    3.48   &    10.9   \\
$40^\circ$  &    0.830    &   0.808      &&   0.88   &    2.31   &    7.23   \\
$50^\circ$  &    0.857    &   0.839      &&   0.60   &    1.56   &    4.90   \\
$65^\circ$  &    0.880    &   0.876      &&   0.29   &    0.76   &    2.38   \\
$70^\circ$  &    0.884    &   0.869      &&   0.25   &    0.65   &    2.05   \\ [6pt]
\hline   \\ [-9pt]
\multicolumn{7}{l}{$^\dag T_2/T_d$ is the temperature ratio required to produce the observed} \\
\multicolumn{7}{l}{\enspace amplitude of the orbital modulation assuming $R_d = 0.9 R_{L1}$.}
\end{tabular}

\end{center}
\end{table}

\begin{table}
\begin{center}
{Table 2: Typical Temperature of the Irradiated\\
          Surface of the Secondary Star} \\ [9pt]
\begin{tabular}{ccc}
\hline\hline
\\ [-9pt]
           &\multicolumn{2}{c}{Temperature (K)} \\ [3pt]
         \cline{2-3}  \\ [-9pt]
Mass Ratio &  $L_X = 10^{37}$ erg s$^{-1}$ & $L_X = 10^{38}$ erg s$^{-1}$ \\ [6pt]
\hline
\\ [-9pt]
    0.3    &          28,000         &         50,000         \\
    0.1    &          22,000         &         40,000         \\
    0.025  &          17,000         &         31,000         \\ [6pt]
\hline  
\end{tabular}
\end{center}
\end{table}

\clearpage

\begin{figure}
\center  \includegraphics[scale=0.65, angle=0]{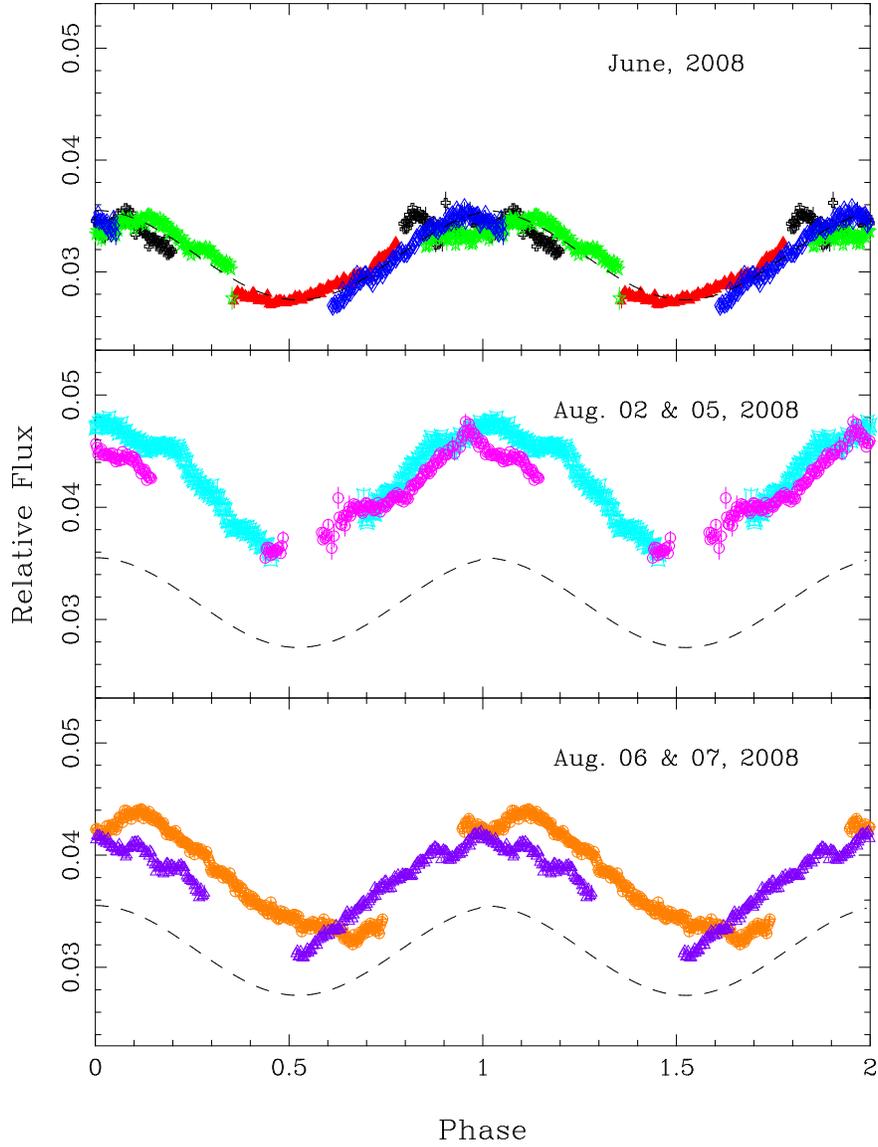}
\caption{The binned light curves of V1408 Aql.  
Each color and symbol combination indicate an individual night. 
{\it Top Panel} -- The fainter state from 2008 June 2--7. 
{\it Middle Panel} -- Brighter state from 2008 August 2 and 5.  
{\it Bottom Panel} -- Transition from the brighter state to fainter 
state on 2008 August 6--7.  
The dashed line is a sine curve fitted to the June data 
(top panel) and plotted in the middle and bottom panels 
for comparison.
Because V1408~Aql was in transition on August~6 and 7, the orbital 
modulation is distorted, causing a spurious shift in the orbital 
phase of maximum light.}
\label{lc}
\end{figure}

\begin{figure}
\center  \includegraphics[scale=0.65, angle=270]{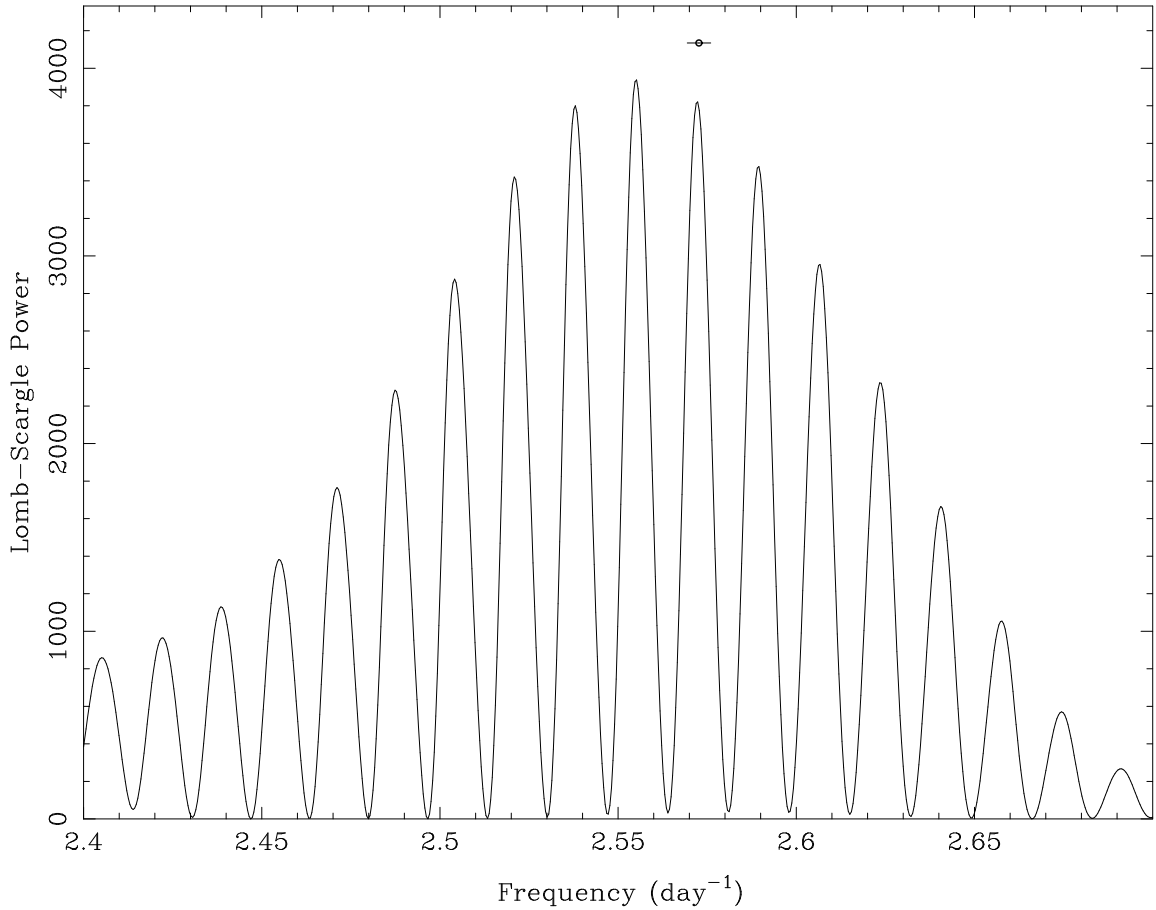}
\caption{The Lomb-Scargle periodogram of the V1408~Aql optical data. 
The multiple peaks are all two-month aliases of each other introduced by 
the two-month separation of our observing runs.
The small circle and error bar mark the $9.329\pm0.011$~h period 
from \citet{tho87}. 
While the highest peak is at a frequency of 2.5552 d$^{-1}$, the peak
at $2.5722\pm0.0001$ d$^{-1}$ or $9.331\pm0.003$~h
agrees with the \citet{tho87} period to within a standard deviation.
It is this period that we adopt as the orbital period. }
\label{lomb}
\end{figure}

\begin{figure}
\center  \includegraphics[scale=0.65, angle=270]{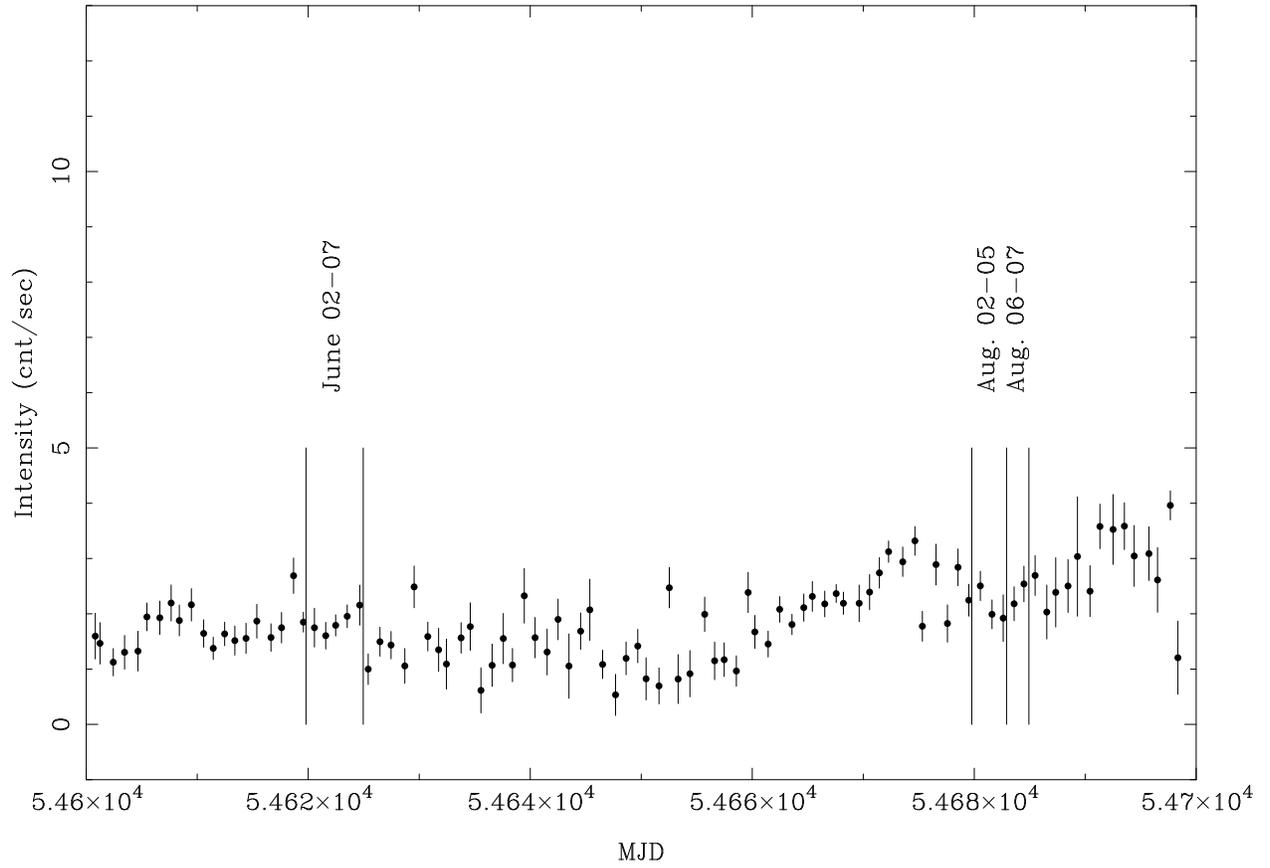}
\caption{The one-day-averaged X-ray light curve from RXTE ASM. 
The times covered to the optical observations in Figure \ref{lc} are 
indicated with the vertical lines.}
\label{xte}
\end{figure}

\begin{figure}
\center  \includegraphics[scale=0.65, angle=270]{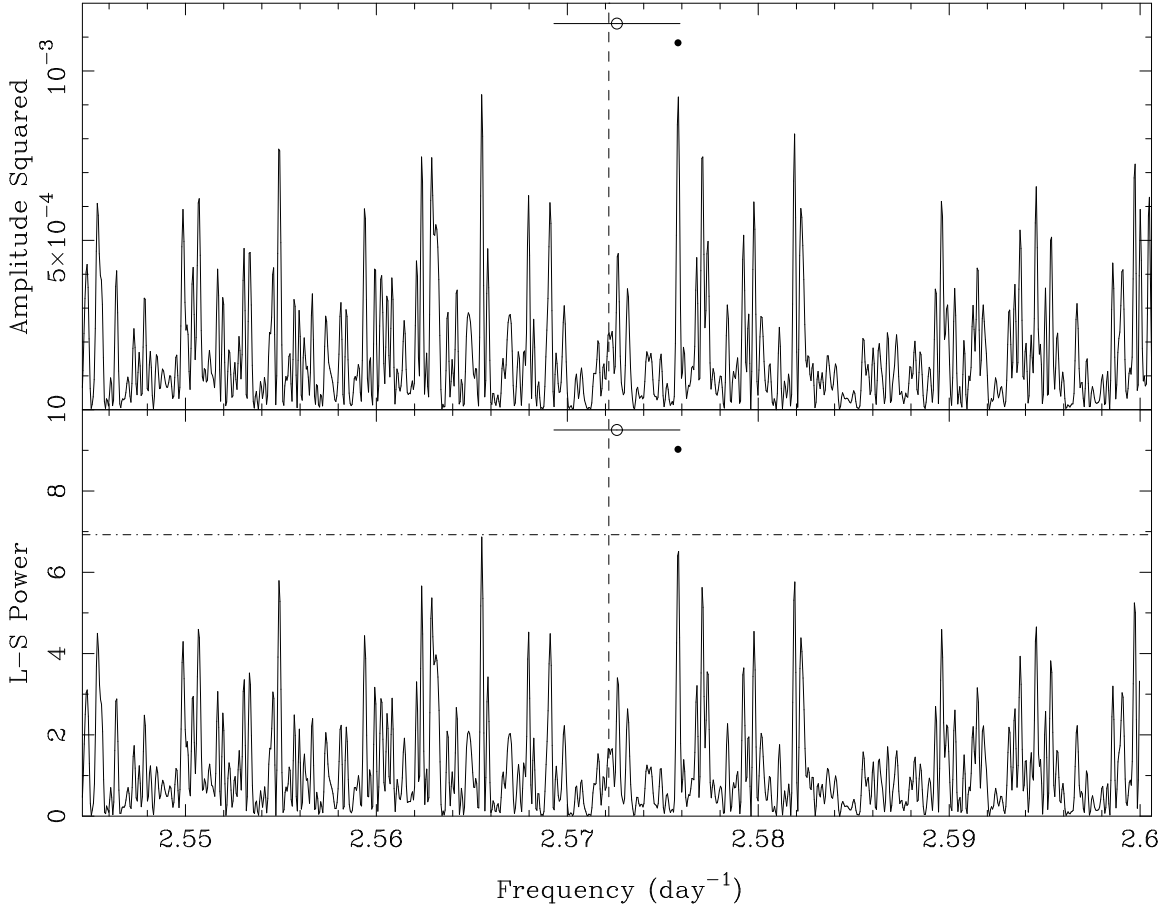}
\caption{The weighted sine curve (top) and weighted 
Lomb-Scargle (bottom) periodograms of the RXTE/ASM X-ray light 
curve of 4U~1957+115.  
The open circle with error bars marks the orbital frequency determined 
by \citet{tho87} and the vertical dashed line marks our refinement 
of the orbital period (equation~\ref{ephem}). 
The dot marks the ``marginal detection'' at $P=9.3175$~h or 
2.5758~d$^{-1}$ found by \citet{lev06}.
The horizontal dot-dash line in the bottom panel indicates the 
99\% false alarm probability. 
All of the peaks are consistent with noise.}
\label{xls1}
\end{figure}

\begin{figure}
\center  \includegraphics[scale=0.65, angle=270]{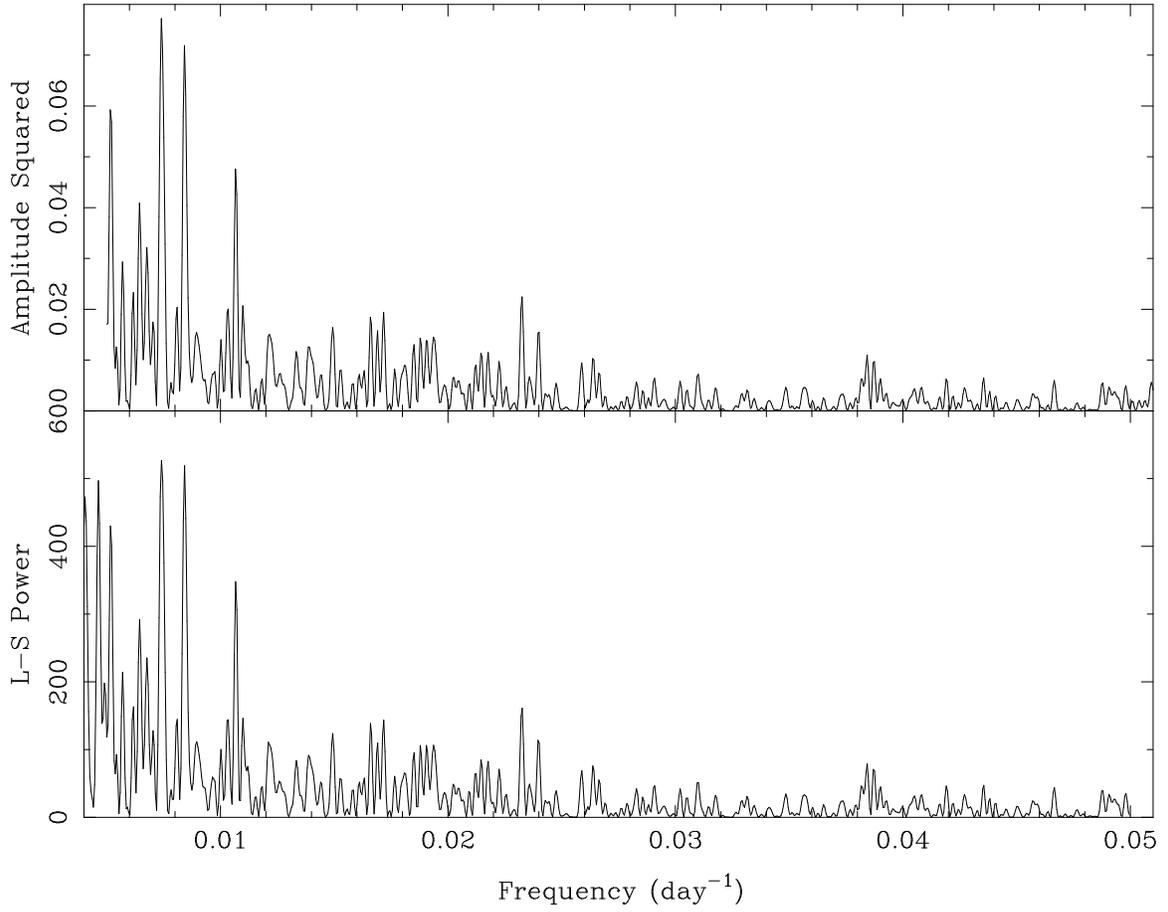}
\caption{The low-frequency portion of the weighted sine curve (top) 
and weighted Lomb-Scargle (bottom) periodograms of the 
RXTE/ASM X-ray light curve of 4U~1957+115.}
\label{xsine}
\end{figure}

\begin{figure}
\center  \includegraphics[scale=0.65, angle=270]{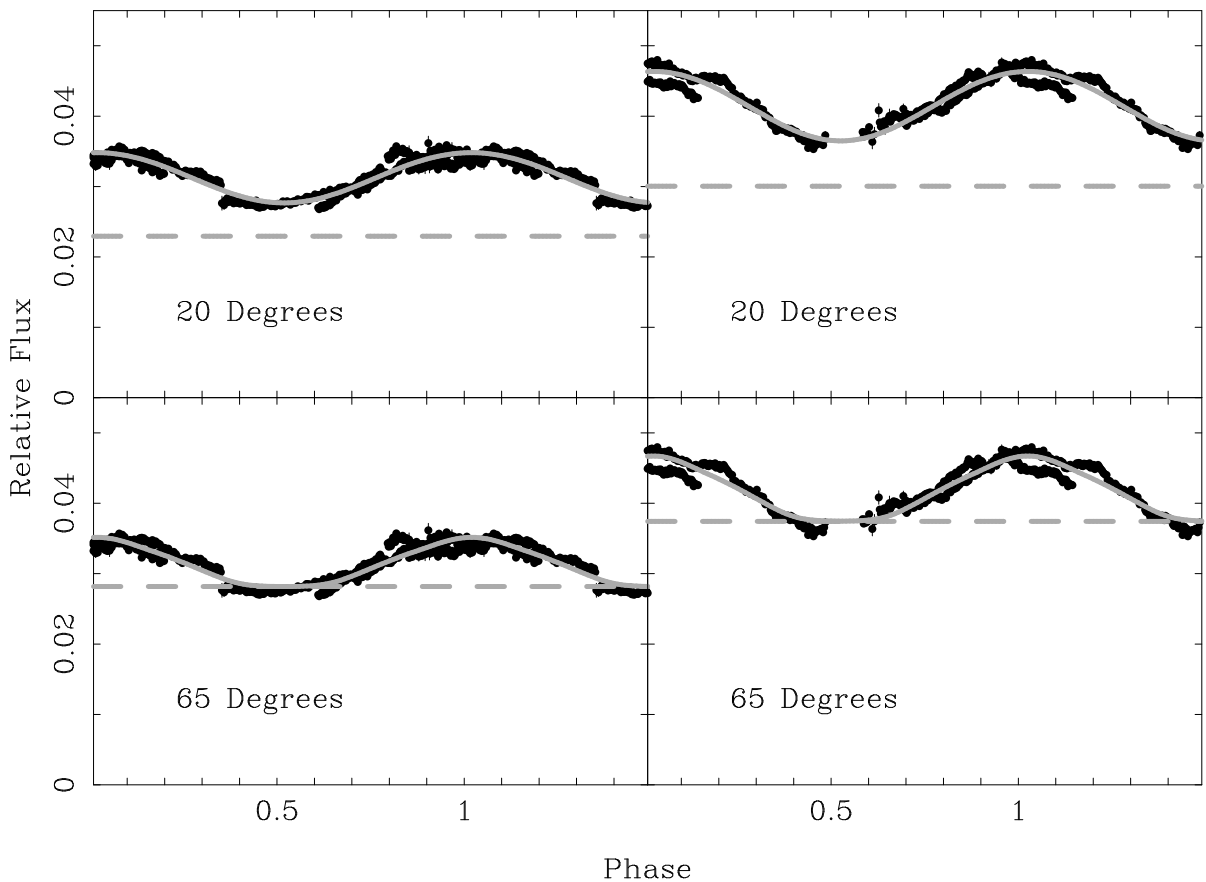}
\caption{The observed orbital light curves of V1408~Aql in 
its low state (left panels) and high state (right panels).
The observed light curves are overlayed with fitted synthetic light
curves calculated for $q = 0.3$.
The orbital inclinations are 
$i = 20^{\circ}$ (top panels) and $i = 65^{\circ}$ (bottom panels).
The dashed line shows the contribution from the accretion disk,
the remaining flux coming from the heated secondary star.}
\label{nslc}
\end{figure}

\begin{figure}
\center  \includegraphics[scale=0.65, angle=270]{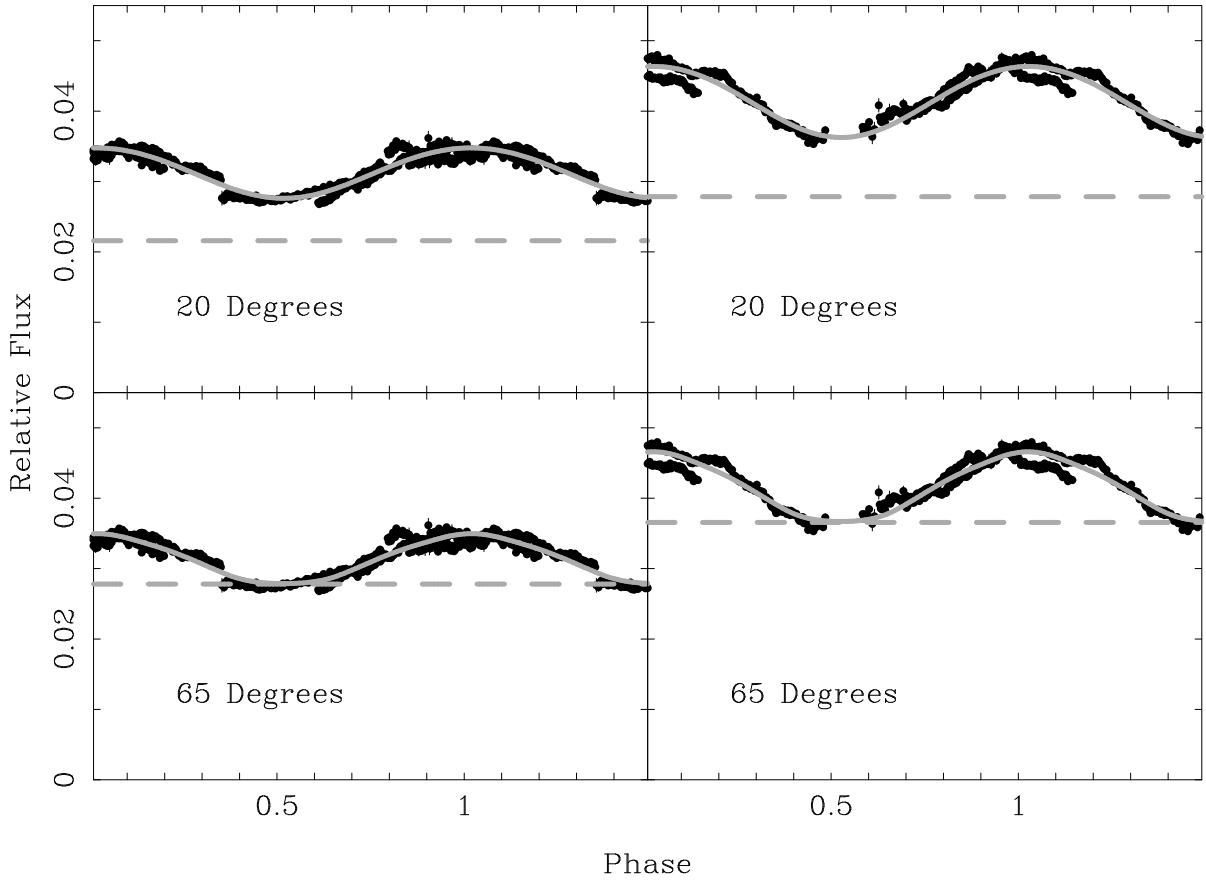}
\caption{Same as Figure \ref{nslc} but for $q=0.025$.}
\label{hmlc}
\end{figure}

\begin{figure}
\center  \includegraphics[scale=0.65, angle=270]{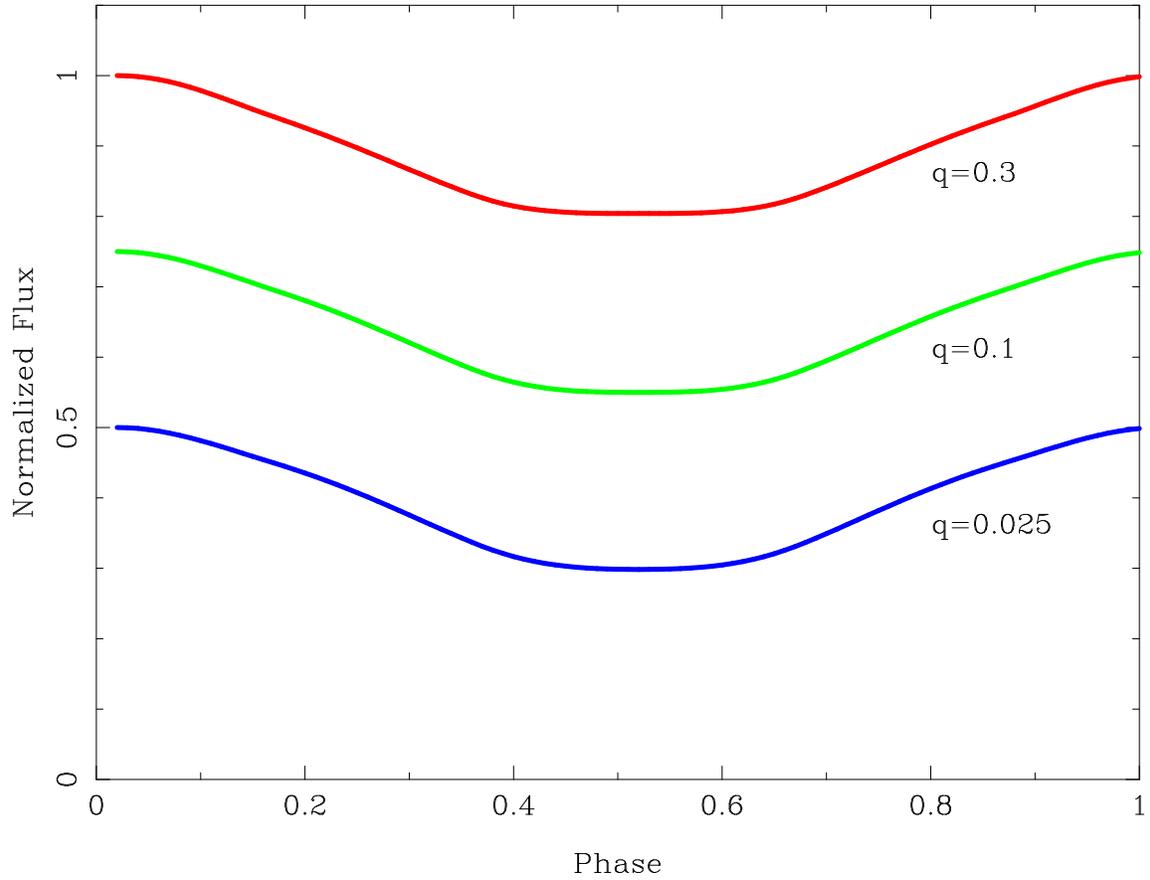}
\caption{Synthetic light curves fitted to the observed light curves
of V1408~Aql in its fainter state.
The orbital inclination is $70^{\circ}$ for each light curve, and
the mass ratios are $q=0.3$ (top, red line), $q=0.1$ (middle, green line),
and $q=0.025$ (bottom, blue line).  
The models have been normalized to 1.0 at maximum and then
offset by 0.25 for clarity.  
At this and higher orbital inclinations the minimum of the light curve 
is flat bottomed.}
\label{70ls}
\end{figure}

\end{document}